# Mechanical and liquid phase exfoliation of cylindrite: a natural van der Waals superlattice with intrinsic magnetic interactions

*Yue Niu[1], Julia Villalva[2], Riccardo Frisenda[3], Gabriel Sanchez-Santolino[3], Luisa Ruiz-González[4], Emilio M. Pérez [2], Mar García-Hernández [3,\*], Enrique Burzurí [2,\*], Andres Castellanos-Gomez [3,\**]

[1]National Center for International Research on Green Optoelectronics & Guangdong Provincial Key Laboratory of Optical Information Materials and Technology, Institute of Electronic Paper Displays, South China Academy of Advanced Optoelectronics, South China Normal University, Guangzhou 510006, China.

[2] IMDEA Nanociencia, C\Faraday 9, Ciudad Universitaria de Cantoblanco, 28049 Madrid, Spain.

[3] Materials Science Factory, Instituto de Ciencia de Materiales de Madrid (ICMM), Consejo Superior de Investigaciones Científicas (CSIC), Sor Juana Inés de la Cruz 3, 28049 Madrid, Spain.

[4] Departamento de Química Inorgánica, Universidad Complutense de Madrid, Madrid, Spain.

Email: MGH marmar@icmm.csic.es, EB enrique.burzuri@imdea.org, ACG andres.castellanos@csic.es

ABSTRACT

We report the isolation of thin flakes of cylindrite, a naturally occurring van der Waals superlattice, by means of mechanical and liquid phase exfoliation. We find that this material is a heavily doped p-type semiconductor with a narrow gap (<0.85 eV) with intrinsic magnetic interactions that are preserved even in the exfoliated nanosheets. Due to its environmental stability and high electrical conductivity, cylindrite can be an interesting alternative to the existing two-dimensional magnetic materials.

The study of van der Waals (vdW) heterostructures and superlattices is undoubtedly among the most active fields within the 2D materials research.[1–6] The capability of fabricating materials with tailored electrical and optical properties by assembling dissimilar 2D materials is extremely appealing.[7–13] Up to now, one of the most widespread approaches to fabricate these vdW heterostructures consists in the stacking of individual layers one-by-one using deterministic placement methods.[10,14–21] Although these methods constitute a powerful route to fabricate stacks of virtually any combination of 2D materials, they suffer from some severe drawbacks such as the presence of interlayer contaminant adsorbates [16,22,23] and the difficulty of precisely controlling the orientation angle between the stacked layers.[24–26] The exfoliation of naturally occurring vdW superlattices formed by alternating layers of two different 2D materials originated by a phase segregation process during their formation has emerged as an alternative method to fabricate thin layers of vdW heterostructures avoiding the above mentioned issues. The sulfosalt mineral franckeite has been recently reported as the first example of exfoliated naturally occurring vdW superlattices.[27–32] Interestingly, the sulfosalt family has other examples of natural vdW superlattices that could be exfoliated as well.

Here we present the first study on mechanically and liquid phase exfoliated (LPE) cylindrite flakes, another mineral member of the sulfosalt family. The isolated flakes are first characterized structurally by TEM and Raman spectroscopy. We then fabricated field effect devices and photodetectors to study their electrical and optical properties. Interestingly, although cylindrite is a superlattice formed by the stacking of two large band gap semiconductors, we find that cylindrite flakes are narrow band gap semiconductors (<0.85 eV). This illustrates very well how a superlattice differs from the trivial sum of the properties of the individual constituents. Finally, we also provide a magnetic characterization of bulk and exfoliated cylinders which display magnetic correlations below 20 K. Cylindrite and other intrinsically magnetic natural heterostructures could therefore be an interesting alternative to expand the exiguous family of the 2D magnetic materials [33–35]. Cylindrite presents the additional advantage of being stable in ambient conditions and conducting and therefore of potential interest for magneto-transport [36,37], in contrast with other insulating magnetic 2D materials.

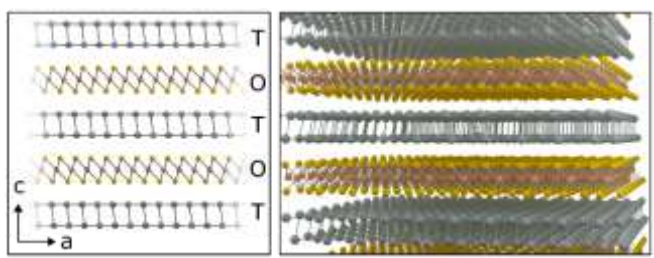

Figure 1. Schematic diagram of the crystal structure of cylindrite, composed of alternating octahedral (O) layers with a $SnS_2$ like structure and tetrahedral (T) layers with a structure similar to the PbS.

Cylindrite belongs to the sulfosalt mineral family and it has an approximate formula $Pb_3Sn_4FeSb_2S_{14}$. Within this family, it is part of a subclass of minerals called misfit compounds that are characterized by a peculiar crystal structure composed of stacks of alternating $SnS_2$-like octahedral (O) and PbS-like pseudo-tetragonal (T) layers.[38–40] This feature is originated by a phase segregation of the two crystalline phases during the rock formation. Interestingly, the growth conditions to reproduce the structure of cylindrite can be replicated in the laboratory to synthesize artificial crystals with cylindrite structure but with user-tailored composition.[41,42] Figure 1 shows a simplified representation of the crystal structure of cylindrite where for the sake of simplicity substitutional Fe, Sb and Sn atoms are not displayed.

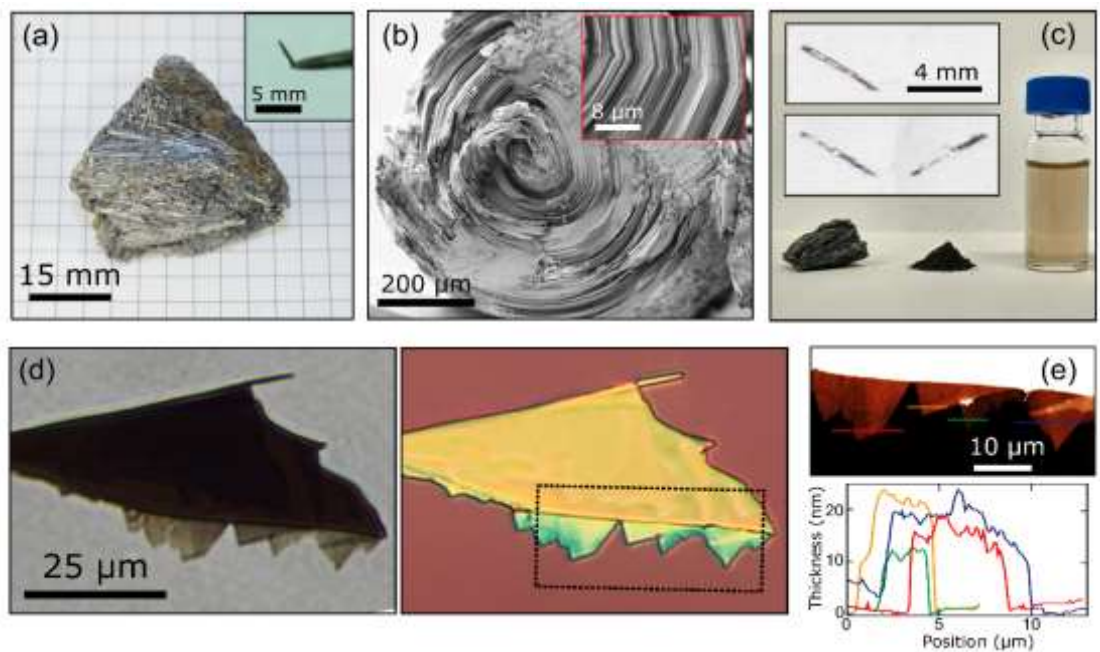

Figure 2. (a) Cylindrite samples: bulk mineral and an individual cylinder extracted from the rock (inset). (b) SEM image showing the base of a single cylindrite cylinder. The inset shows an amplified region around the revolution axis of the cylinder. The layered structure and the kinks source of the cylindrical shape are clearly observed. (c) Cylindrite samples. Left: bulk mineral; middle: powder material obtained after grinding several cylinders in an agate mortar; right: suspension of exfoliated material prepared by sonication of a 1 mg·ml$^{-1}$ powder dispersion in iPrOH. (inset in c) Optical images of macroscopic cylindrite flakes peeled off from the cylindrite cylinder with Scotch tape. (d) Optical images of a cylindrite flake on a Gel-Film stamp before its transfer (left) and after transferring it onto a SiO$_2$/Si substrate (right). (e) AFM topographic image of the area highlighted with a dashed square in (d). The corresponding thicknesses of different cylindrite flakes.

Figure 2a shows an optical image of cylindrite mineral displaying its characteristic appearance with clusters of small diameter (< 1 mm) and long (~ 3 – 5 mm) cylinders. The individual cylinders can be easily extracted from the rock (the inset of Figure 2a shows one of these cylinders held at the tip of lab tweezers). Since the cylinders are formed by concentric layers of material with vdW mediated interlayer interactions (see Figure 2b), they can be easily exfoliated with adhesive tape (mechanical exfoliation) or by LPE (Figure 2c). Figure 2c shows the different steps to carry out the LPE of cylindrite[32,43–45]: 1) several cylindrite cylinders were grinded carefully in an agate mortar until a black powder was obtained, 2) the powder was dispersed in iPrOH at a 1 mg·ml$^{-1}$ concentration, 3) this dispersion was sonicated for 1 h in an ultrasonic bath kept at 20 ºC, 4) the as-prepared suspension was centrifuged (990 g, 30 min, 20 °C, Beckman Coulter Allegra X-15R, FX6100 rotor, radius 9.8 cm) in order to eliminate thicker (and thus heavier) flakes. After this process, the supernatant was collected carefully to obtain a pale orange-colored exfoliated cylindrite suspension (see Figure 2c). The suspension precipitates at room temperature within one week and the nanosheets can be easily re-dispersed by sonication. Alternatively to the LPE method, cylindrite flakes can be also isolated by mechanical exfoliation. The inset in Figure 2c shows an optical image of macroscopic flakes peeled off from the surface of a cylindrite cylinder with Scotch tape exfoliation. The flakes can then be transferred onto a Gel-Film (WF x4, by Gel-Pak®) substrate by placing the tape containing the flakes in contact to the Gel-Film surface and peeling it off fast (Figure 2d). From the Gel-Film, the cylindrite flakes can be transferred to an arbitrary substrate by an all dry deterministic transfer method.[15] Figure 2d also shows an optical image of the same flake after being transferred onto a SiO$_2$/Si substrate. The topography of the cylindrite flakes have been studied with atomic force microscopy (AFM) to determine their thickness (Figure 2e).

We characterized the structure of the mechanically and LPE cylindrite flakes by transmission electron microscopy (TEM). Figure 3a shows a low magnification TEM image of a mechanically exfoliated cylindrite flake transferred onto a holey silicon nitride TEM grid. A striped pattern of darker and lighter areas can be seen in the low magnification image. This pattern arises from the rippled structure of the cylindrite structure that is due to a deformation of the lattice that occurs to force the commensuration of the two incommensurate lattices. HRTEM images, as shown in Figure 3b, show the characteristic stacking of the pseudo-tetragonal (T) and octahedral (O) layers of the cylindrite.[14] The fringes are approximately spaced 3.56 nm along the c direction. Figure 3c shows the indexed fast Fourier transform (FFT) of the image in Figure 3b in which the reflections corresponding to the T and O layers are marked with yellow and red circles respectively. Figure 3d and 3e shows the corresponding low TEM and HRTEM images of a LPE cylindrite flake showing the same structural features as the mechanically exfoliated flakes. In addition, the energy-dispersive X-ray spectroscopy (EDS) shown in Figure 3f unveils the presence of Fe in the nanosheets, expected in cylindrite also after exfoliation.

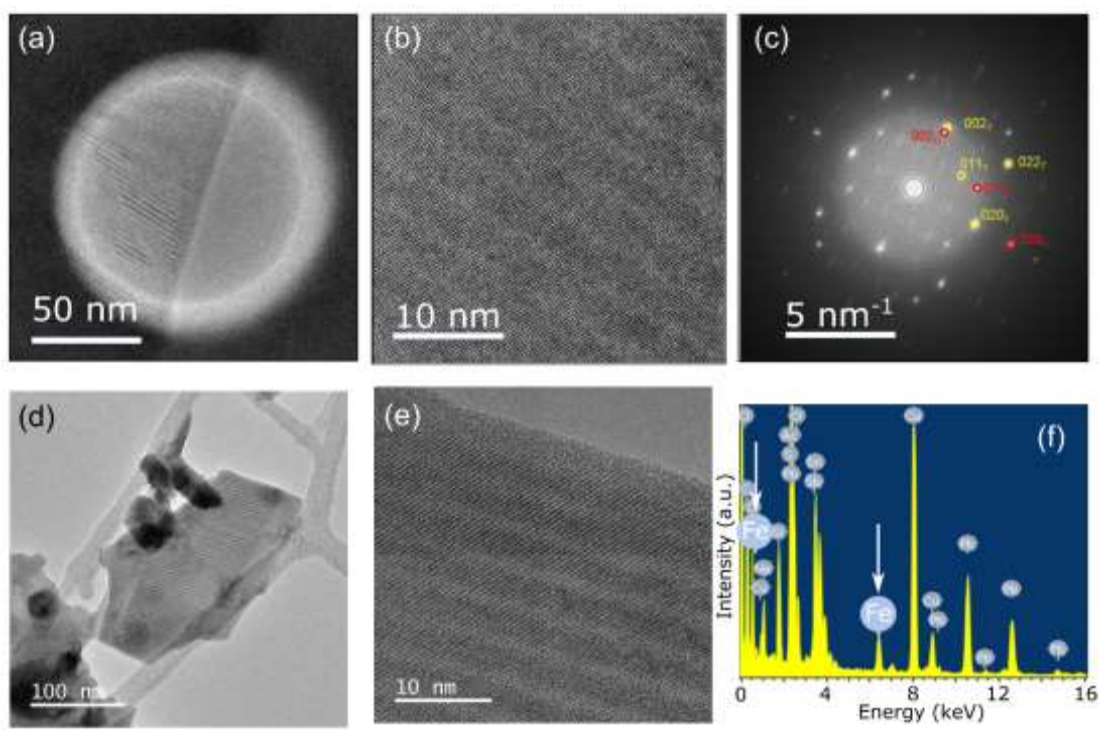

Figure 3 (a) A low magnification TEM image of a mechanically exfoliated cylindrite flake. (b) HRTEM image of an ultrathin mechanically exfoliated cylindrite layer. (c) FFT of b, consistent with the stacking of the pseudo-tetragonal (T: red) and octahedral (O: yellow) layers. (d) Low magnification TEM image of a LPE cylindrite flake. (e) HRTEM image with atomic resolution of a LPE cylindrite flake. (f) EDS spectrum taken on a single ultrathin cylindrite nanosheet. The arrows highlight the presence of Fe in the flake.

In order to characterize the electrical and optical properties of cylindrite we have fabricated simple field effect devices and photodetectors. Figure 4a shows the current *vs*. voltage (*IV*) characteristic of a cylindrite device (see inset). The corresponding source-drain current vs. gate voltage trace is displayed in the bottom inset in Figure 4a, showing a marked p-type doping without reaching the OFF state of the device (the conductance decreases monotonically upon increasing the gate voltage without vanishing).

Interestingly, cylindrite shows a sizeable photoresponse even for the illumination wavelength of 1550 nm, indicating that its bandgap is <0.85 eV. Figure 4b shows the photocurrent ($I_{ph}$) generated upon pulsed illumination with a wavelength of 1550 nm at different power densities ranging from 0.10 to 0.51 W/cm$^2$. The relationship between the photocurrent and the light intensity is shown in Figure 4c. It can be fitted well with a power law: $I_{ph} \sim P^{\Theta}$, where exponent $\Theta$ determines the photo response to light intensity. By fitting the curve, $\Theta = 0.82$ is obtained, which indicates the presence of charge carrier traps in the flake. From this plot one can extract the power dependence of the responsivity $R=I_{ph}/(P_d*A)$, where $I_{ph}$ is the photocurrent, $P_d$ is the power density of the illumination and $A$ is the active area of the device. The maximum responsivity we obtain at 1550 nm is 0.83 mA/W, which is comparable with other reported narrow bandgap two-dimensional photodetector materials such as black phosphorus.[46–48]

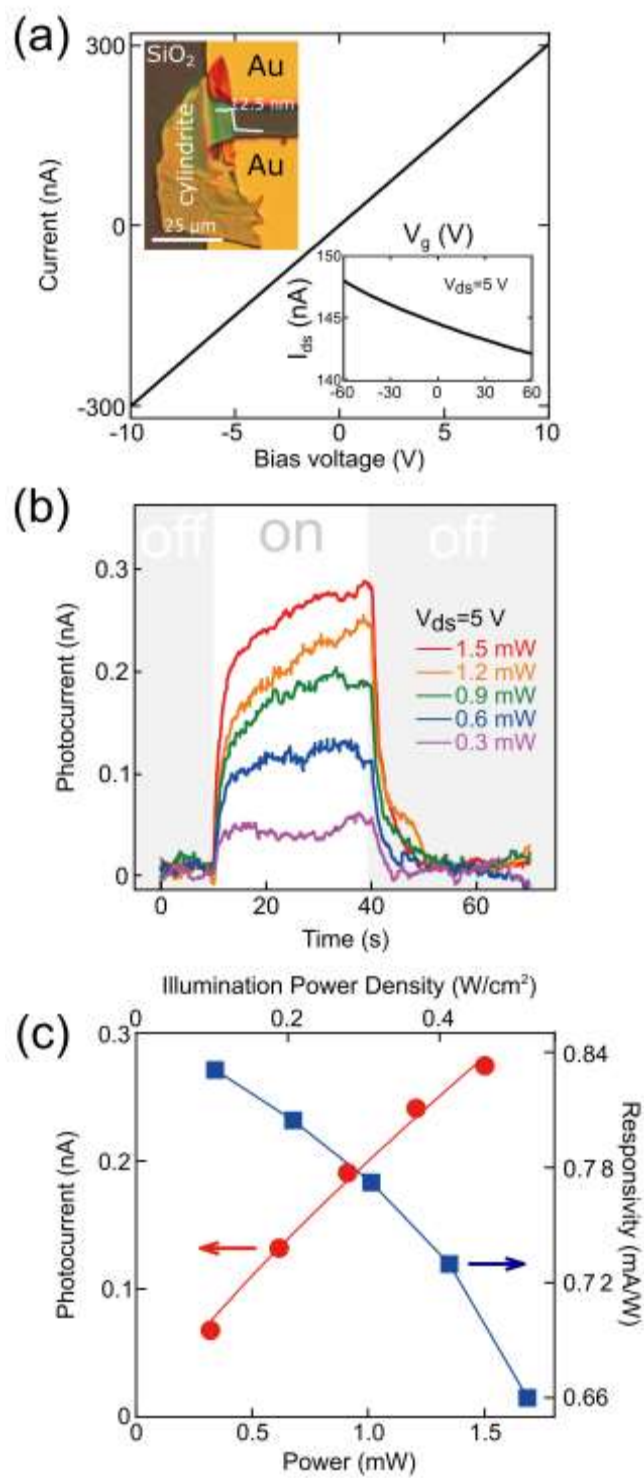

Figure 4 (a) Current vs. bias voltage characteristics of the a cylindrite device in dark. (Top inset) Optical microscopy image of the fabricated cylindrite device. (Bottom inset) Transfer characteristics ($I$–$V_g$) in dark at 5 V of bias. (b) Photocurrent versus time recorded at 5 V while switching on and off the illumination at 1550 nm with increasingly high power. (c) Power dependence of the photocurrent (red) and responsivity (blue) with 1550 nm and a bias voltage of 5 V.

In order to study the magnetic properties of cylindrite, we first measured the magnetism of bulky cylinders mechanically extracted from the mineral rock and we compared them to those of dispersed thin nanosheets stemming from LPE. Note that the very same cylinders isolated for the measurement in bulk are used to produce the LPE dispersion. This is important to ensure strictly the same chemical composition in both samples, as a natural mineral is the source to prepare them. The concentration of our suspension, 1 mg·ml$^{-1}$, allows a 'macroscopic' sample of exfoliated thin flakes of cylindrite dispersed in the selected solvent. In spite of its potential interest to explore low-dimensional magnetism in layered and nanoscale materials, LPE has been scarcely used to explore the magnetism of other bidimensional materials different form functionalized graphene. Indeed, to the best of our knowledge, only $WS_2$ dried precipitates have been explored after exfoliation in the liquid phase.[49]

The magnetism of the bulk cylindrite and LPE nanosheets has been studied using a SQUID magnetometer from Quantum Design equipped with a 5 Tesla coil. Figures 5(a,b) show the temperature dependence of the magnetization measured in a set of three cylindrite bulk cylinders with similar aspect ratio and packed together with their long axes aligned parallel. In a zero-field cooled (ZFC) measurement, the sample is first cooled down in the absence of a magnetic field. The magnetization is thereafter measured while ramping the temperature up under an applied magnetic field $H$ =1000 Oe parallel (Figure 5a) or perpendicular (Figure 5b) to the cylinders' axial direction.

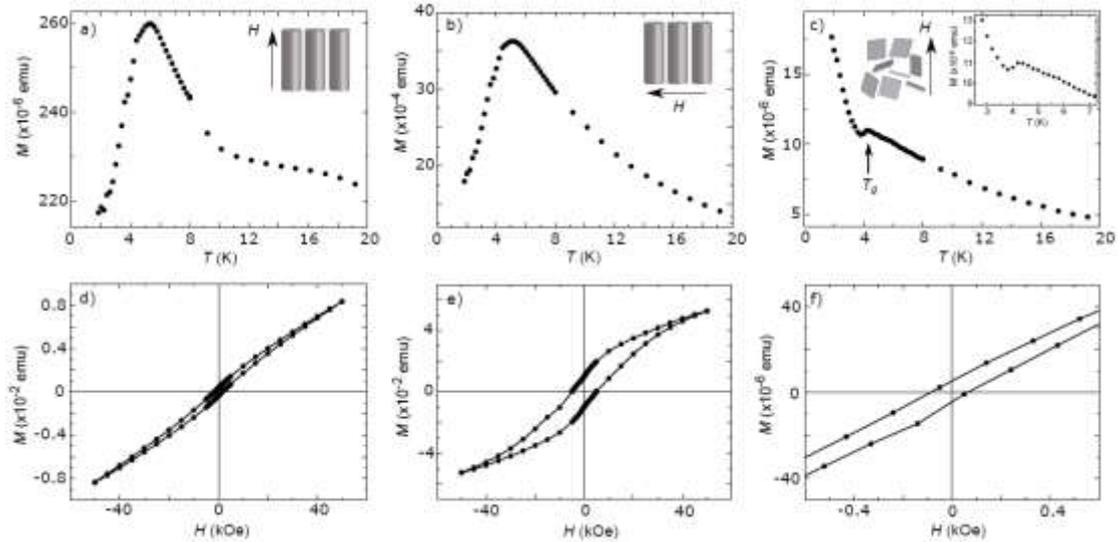

Figure 5. (a,b) Zero field cooled (ZFC) magnetization measured as a function of temperature under $H$ =1000 Oe applied parallel (a) or perpendicular (b) to the axial direction of the cylinders. (c) ZFC magnetization measured in a suspension of cylindrite nanosheets obtained by LPE. (d-f) Magnetization measured as a function of the magnetic field at $T$ =1.8 K for the three corresponding configurations.

The magnetization monotonically increases while decreasing temperature down to $T_k$ = 10 K. Below $T_k$ = 10 K a sharp increment of $M$ is triggered that peaks at $T_{g,\|}$ = 5.3 K when the field is parallel to the cylinders' axis. A similar peak at $T_{g,\perp}$ = 5.1 K but with a less pronounced kink is observed for the perpendicular orientation of $H$. This anomaly in the magnetization can be attributed to a slowing down of the spin dynamics in magnetically disordered systems, the hallmark of spin glass-like behavior. It has been reported for cylindrite at similar temperatures.[50] The magnetic frustration could be originated by the random distribution of Fe in the magnetically diluted cylindrite lattice. The spin glass-like interactions manifest also in the magnetization measured as a function of the magnetic field, shown in Figures 5(d,e). A small, elongated sigmoidal-like hysteresis, typical of spin glass-like systems, opens with a coercive field $H_{c,\|}$ ≈ 1.58 kOe for $H$ parallel to the cylinders' long axis. This value reaches up to $H_{c,\perp}$ ≈ 5.05 kOe for the perpendicular configuration (Figure 5e). The two orders of magnitude difference in ZFC magnetization and the different $H_c$ between parallel and perpendicular configurations is originated by the magnetic anisotropy experienced by the Fe ions in the lattice.[51]

After the bulk measurements are completed, the same three cylinders are used to obtain cylindrite nanosheets dispersed in iPrOH by LPE as detailed before. A droplet of the dispersion containing roughly hundreds of nanoflakes is mounted in a sealed plastic diamagnetic capsule and thereafter frozen for SQUID measurements. Figure 5c shows the resulting ZFC magnetization curve measured as a function of temperature. The background contribution to the magnetic signal of solvent and capsule is corrected by subtracting a control measurement on a capsule containing only iPrOH. Again, a small peak is observed at around $T_g = 4.3$ K, close to bulk $T_g$, followed by a sharper increment of $M$ down to the lowest measured temperatures. The small peak seems related to the spin-glass-like transition observed in bulk while the latter increment in $M$ may be associated to an increase of isolated Fe ions resulting from the LPE process behaving as a paramagnet, although an underestimated contribution to the background cannot be ruled out. In addition, the *M versus H* measurement (Figure 5f) shows a small hysteresis with reduced $H_c \approx 76$ Oe compared to bulk signaling somehow a "magnetic softening" of the exfoliated flakes. This points to a decrease of harder magnetic interactions that, in turn, could be explained by the lack of interlayer interactions in the two-dimensional flakes, as compared to the tridimensional material. The persistence of a peak in the ZFC magnetization and the small hysteresis indicate therefore that the magnetic interactions are preserved in the exfoliated nanosheets. The reduced $H_c$ and the slightly decrease of $T_g$ could be also ascribed to the decrease in the effective size of the sample during the LPE synthesis, not only in the thickness but also in the lateral dimension. This behavior has been reported in other magnetic materials when their size is reduced to the nano/micro-scale [52–54] or their magnetic lattice is repeatedly diluted [55] and therefore the number of interacting spins is dramatically reduced.

In summary we demonstrated the isolation of thin flakes of cylindrite by mechanical and liquid-phase exfoliation of cylindrite bulk crystal and we provided a first structural, magnetic, electrical and optical characterization. In particular, we showed that cylindrite nanosheets are highly doped p-type semiconductors with a narrow bandgap. These features, in combination with its inherent environmental stability, make cylindrite an alternative material to black phosphorus. We also found that cylindrite presents magnetic interactions that are preserved even in the exfoliated nanosheets. In contrast with other reported magnetic 2D materials, cylindrite has the advantage of being stable in ambient conditions and it is electrically conducting. Moreover, we demonstrated that liquid-phase exfoliation of magnetic layered materials is a powerful tool to explore the magnetism of layered heterostructures at low-dimensionalities and at the nano-scale.

**Materials and Methods**

*Materials*
The cylindrite crystal source used in this work is a natural mineral from San José mine (Oruro, Bolivia).

*Scanning electron microscopy (SEM)*
The SEM images are recorded by a secondary electrons detector mounted in a Carl Zeiss AURIGA Scanning Electron Microscope. The acceleration voltage is 2 kV and the working distance is 8 mm.

*Transmission electron microscopy (TEM)*
High resolution transition electron microscopy (HRTEM) observations were carried out in an aberration-corrected JEOL JEM-GRAND ARM300cF operated at 120kV and equipped with a cold field emission gun and a Gatan OneView camera.

*Atomic force microscopy (AFM)*
Atomic force microscopy images were carried out in a Nanotec Cervantes AFM operated in tapping mode with cantilevers with a spring constant of 40 N/m and 300 kHz of resonance frequency.

*Optical microscopy images*

Optical microscopy images have been acquired with an AM Scope BA310 MET-T upright metallurgical microscope equipped with an AM Scope mu1803 camera with 18 megapixels.

*Transport measurements*

Transport measurements have been carried out in a homebuilt probe station equipped with a source measuring unit (2450 Keithley) and a probe with a multimode optical fiber illuminator. A 1550 nm fiber coupled laser (KLS1550 Thorlabs) was used to measure the photoresponse of cylindrite in the short-wavelength infrared regime.

*Magnetic measurements*

The magnetization in the samples is measured in a Quantum Design SQUID magnetometer mounted in a low temperature cryostat. The bulk cylinders are stacked together with kapton tape in a diamagnetic capsule.

**ACKNOWLEDGMENT**

YN acknowledges the grant from the China Scholarship Council (File NO. 201506120102). ACG acknowledges funding from the European Research Council (ERC) under the European Union's Horizon 2020 research and innovation programme (grant agreement n° 755655, ERC-StG 2017 project 2D-TOPSENSE) and from the EU Graphene Flagship funding (Grant Graphene Core 2, 785219). RF acknowledges support from the Netherlands Organisation for Scientific Research (NWO) through the research program Rubicon with project number 680-50-1515. EB acknowledges funding from the European Commission under the Marie Sklodowska-Curie programme (MSCA-IF 746579) and the Comunidad de Madrid Atracción del Talento Programme (2017-T1/IND-5562). E.M.P. acknowledges funding from the European Research Council (ERC-StG-MINT 307609), the Ministerio de Economía y Competitividad (CTQ2014-60541-P, CTQ2017-86060-P), and the Comunidad de Madrid (MAD2D-CM program S2013/MIT-3007) IMDEA Nanociencia acknowledges support from the 'Severo Ochoa' Programme for Centres of Excellence in R&D (MINECO, Grant SEV-2016-0686). We thank the National Centre for Electron Microscopy (ICTS-CNME, Universidad Complutense) for electron microscopy facilities. G.S.S was supported by the Juan de la Cierva program FJCI-2015-25427 (MINECO-Spain).

# Supporting Information:

# Mechanical and liquid phase exfoliation of cylindrite: a natural van der Waals superlattice with intrinsic magnetic interactions

*Yue Niu[1], Julia Villalva[2], Riccardo Frisenda[3], Gabriel Sanchez-Santolino[3], Luisa Ruiz-González[4], Emilio M. Pérez [2], Mar García-Hernández [3,*], Enrique Burzurí [2,*], Andres Castellanos-Gomez [3,*]*

[1]*National Center for International Research on Green Optoelectronics & Guangdong Provincial Key Laboratory of Optical Information Materials and Technology, Institute of Electronic Paper Displays, South China Academy of Advanced Optoelectronics, South China Normal University, Guangzhou 510006, China.*

[2] *IMDEA Nanociencia, C\Faraday 9, Ciudad Universitaria de Cantoblanco, 28049 Madrid, Spain.*

[3] *Materials Science Factory, Instituto de Ciencia de Materiales de Madrid (ICMM), Consejo Superior de Investigaciones Científicas (CSIC), Sor Juana Inés de la Cruz 3, 28049 Madrid, Spain.*

[4] *Departamento de Química Inorgánica, Universidad Complutense de Madrid, Madrid, Spain.*

Email: MGH marmar@icmm.csic.es, EB enrique.burzuri@imdea.org, ACG andres.castellanos@csic.es

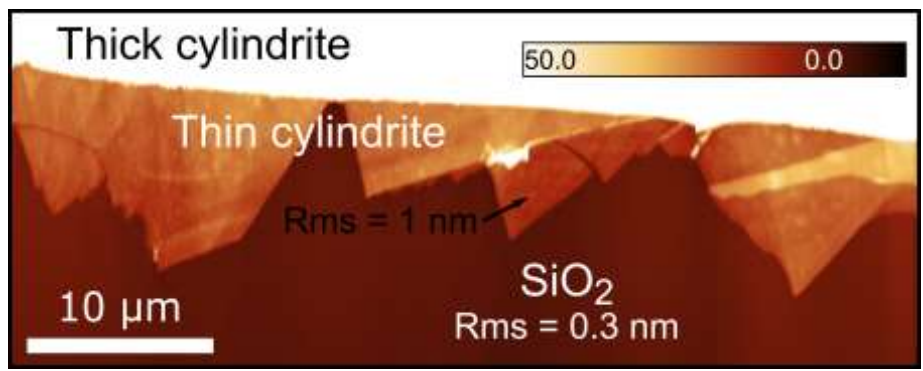

Figure S1. Detail of the AFM topography image displayed in Figure 2 in the main text. Information about the roughness have been included in the figure.

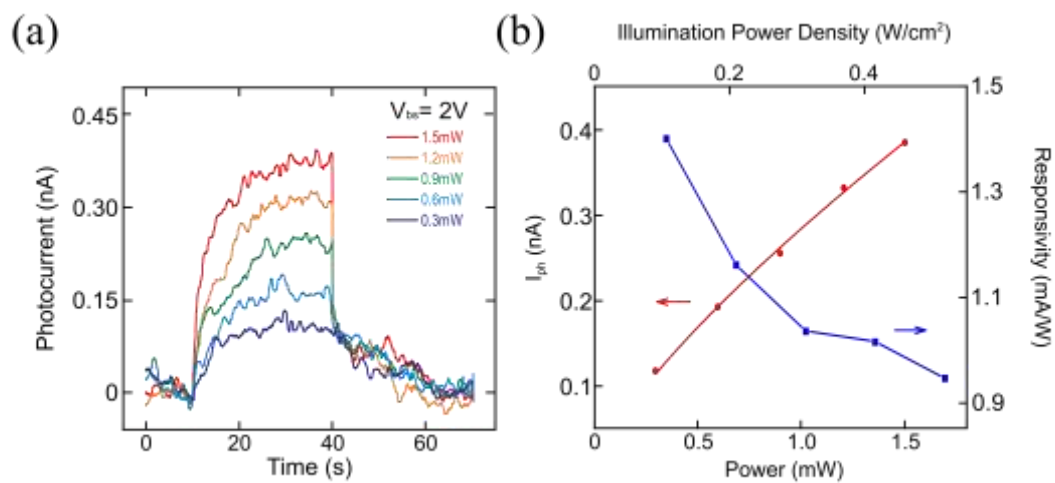

Figure S2. (a) Photocurrent versus time recorded at 2 V while switching on and off the illumination at 1050 nm with increasingly high power. (c) Power dependence of the photocurrent (red) and responsivity (blue) with 1050 nm and a bias voltage of 2 V.